# Comments to Marvel Fusions Mixed Fuels Reactor Concept

K. Lackner, R. Burhenn, S. Fietz, A. v. Müller

Max Planck Institute for Plasma Physics, 85748 Garching, Germany

**Abstract:**
Nanostructured solid boron-hydrogen compounds have been suggested as target and fuel for laser fusion, offering improved laser-plasma coupling, avoiding cryogenic fuel handling and fuel pre-compression and ultimately allowing a transit from DT- to aneutronic pB- fusion power production. We describe the scaling of the different energy loss channels (α-particle escape, bremsstrahlung, hydrodynamic expansion work, electron heat conduction) with mixed fuel composition using partial inverse gains (Q's) which allow a simple superposition of losses. This highlights in particular the negative synergy between these loss-channels for such mixed fuels: the dominance of bremsstrahlung over fusion power at low temperatures forces a shift of operation to higher ones, where the plasma gets more transparent to α-particles, and hydrodynamic and heat conduction losses increase strongly. The use of mixed fuels therefore does not eliminate the need for strong precompression of the fuel: in fact, it renders achieving burning plasma conditions much more difficult, if not impossible. A recent suggestion to use tamping of the fuel by cladding with a heavy metal would only reduce hydrodynamic expansion losses significantly if the cladding could cover most of the fuel surface, in competition with access to laser radiation. But even if tamping were perfect, this would not reduce the remaining three loss channels - in fact it would have a negative effect on burn propagation, as the escaping energy would not heat surrounding fuel, but only the cladding material.

## 1. Introduction

Inertial fusion initiated by compression and heating of *DT* pellets through Laser irradiation has reached important milestones after a targeted 50 years research and development program [1]. This was achieved by carefully programmed laser pulses and layered fuel structures so as to attain simultaneously several hundred-fold solid ice density and ignition temperatures.

Marvel fusion scientists suggest to avoid the need for precompression of the target and of a cryogenic environment by efficient coupling of ultrafast laser beams to nano-structured solid-state targets [2, 3, 4, 5, 6]. Ultimate aim is aneutronic fusion of proton-boron (*pB*) [7], but a fuel mixture provided by suitable boron-hydrogen compounds is considered at the present stage. Self-heating and high target gain is a generally accepted prerequisite for energy production through laser fusion [8]. This need would persist even for schemes with very high efficiency in the conversion of laser into thermal plasma energy, in particular if femto-second lasers with multi-MJ energy are required for initiating fusion reactions. In the following we analyze modifications to the usual ignition requirements arising from the use of such mixed fuels, discuss and correct analytic estimates given by Marvel authors in previous notes, and address their latest proposal concerning tamping of fuel expansion by high mass density cladding [9]. Finally, we address remarks made by Ruhl and Korn (abbreviated in the following by RK) [10] regarding our previous criticism of their proposals. We do not address the laser-matter interaction aspects of their proposal but start from their assumed absorbed laser energy $E_i$ in the hot spot.

## 2. Energy loss channels and dependence on fuel composition

For thermonuclear burn and high energy gain, four necessary criteria have to be satisfied to ensure that the fusion power carried by charged particles ($W_\alpha$) exceeds all energy losses from the hot spot. They are extensively treated in the textbook by Atzeni and Meyer-ter-Vehn (abbreviated in the following as AMtV) [11], but only for simple fuel compositions, primarily $DT$.

These primary loss channels - using the nomenclature of AMtV - are
(a) the energy of escaping charged fusion products ($\alpha$-particles in both cases of $DT$ and $pB$ reactions): $W_{esc} = W_\alpha - W_{dep}$
(b) (bremsstrahlung) radiation: $W_{bre}$
(c) mechanical expansion work of the fluid: $W_m$
(d) electron heat conduction: $W_e$.

We will discuss here the scaling of these losses with fuel composition, starting with the expressions given in AMtV (sections 4.1.). We will measure the losses by the ratio of their respective power to the total produced fusion power in charged particles ($W_\alpha$) and define these quantities as inverse partial $Q's$ :

$$\frac{1}{Q_{esc}} = \frac{(W_\alpha - W_{dep})}{W_\alpha}, \frac{1}{Q_{bre}} = \frac{W_{bre}}{W_\alpha}, \frac{1}{Q_m} = \frac{W_m}{W_\alpha}, \frac{1}{Q_e} = \frac{W_e}{W_\alpha}. \qquad \ldots (1)$$

The advantage of these expressions is that they are simply additive,

$$\frac{1}{Q_\alpha} = \frac{1}{Q_{esc}} + \frac{1}{Q_{bre}} + \frac{1}{Q_m} + \frac{1}{Q_e}, \qquad \ldots (2)$$

and that the hot-spot self-heating condition can be written as

$$\frac{(W_\alpha - W_{dep}) + W_{bre} + W_m + W_e}{W_\alpha} = \frac{1}{Q_\alpha} \leq 1 \qquad \ldots (3)$$

corresponding to satisfaction of equ.4.18 in AMtV. For ease of comparison with AMtV and the latest notes of RK we use cgs units, with the exception of single particle energies and temperatures which we express in keV.

Like AMtV we use for the following estimates $T_e = T_i$, which has been accepted as a good assumption in the DT fusion regime[i] also in the most recent Marvel publication [9], and verified there also by results of simulations.

We write down expressions allowing for the simultaneous presence of 5 types of atoms: protons (p), deuterons (D), tritons (T), boron (B) and an additional atom possibly needed to form an actually existing solid material. For a plasma target nearly arbitrary fuel compositions are possible. The constraint to non-cryogenic solid targets restricts this, however, to certain

---
[i] in the regime of burn of pB, treated, e.g., by Putvinski, slowing down on ions dominates, and $T_e < T_i$ will generally hold

boron composites with the largest possible hydrogen content. For numerical examples we consider 4 different compositions of atom densities $n_i$ and, as reference, the DT case:

DT:     $n_D = n_T$ (as reference case, treated extensively in AMtV).

The three boron-hydrogen composites considered by us include two proposed at different instances by RK:

MIX1: $n_D = n_T = n_p = n_B$ (as the reference case of RK in ref. [5])
MIX2: $0.6n_D = 0.6n_T = 0.6n_p = n_B$ (forming the basis of recent papers of RK [6],[9]).

As it turns out that MIX1 does not satisfy the bremsstrahlung criterium (and MIX2 likely will never exist)[ii], we consider also lithium-borohydride (LiBH4) which - according to our search - is the only solid[iii] which could do so, but only under the condition that no hydrogen spots are used for protons:

LiB:    $n_D = n_T = 2n_B = 2n_{Li}$ .

As *p-B* reactions are negligible in the regime where attainment of self-heating conditions in mixed fuels can be hoped-for, (for $T_i \leq 100\ keV$ their contribution to $\alpha$-power production is less than 1 % - see Putvinski [12]), this appears well justified.

For comparing different fuel compositions, the question of density normalization arises. As in the region of interest for burn initialization considered by RK only the *DT* reaction contributes significantly to fusion power production, we normalize results for mass densities and energies to the partial mass density of $DT$ defined as $\rho_{DT} = (2n_D + 3n_T)m_p$ with the proton mass $m_p$. This allows direct comparison with textbook (AMtV) results for pure $DT$, and is also in line with the latest usage by RK.

2.1. Energy loss via escaping $\alpha$ - particles

Initiating and sustaining thermonuclear burn requires the deposition of a major part of the fusion energy released in charged particles within the hot spot region. For an optimal, spherical target, an expression for the absorbed (and hence also for the escaping) energy fraction is given by AMtV in terms of the ratio of the hot-spot radius ($R_h$) to the slowing down length (equs. 4.6-4.8) in *DT* plasmas. It is based on Coulomb collisions with electrons, and can be easily extended to a general fuel mix. Here, a significant benefit of mixed fuels - at given *DT* density - derives from the enhanced electron number and is shown in Fig. 1 as the dependence of $1/Q_{esc}$ on $T_h^{3/2}/(\varrho_{DT,h} R_h)$. For the isochoric reference case of AMtV (their equs. 4.23, 4.24), with $\varrho_{DT,h} R_h = 0.5\, gcm^{-2}$ and $T_h = 12\ keV$ these losses amount to less

---

[ii] The first compound (MIX1) proposed by RK exists, but is gaseous (diborane); MIX2 as a solid appears unfeasible and would be sensational due to its high gravimetric hydrogen content (> 200kg/m$^{-3}$ for "light" hydrogen).

[iii] ammonia borane NH$_3$BH$_3$, which is also a solid, has a slightly higher gravimetric hydrogen density and is closest to the nominal MIX1 in composition, does not satisfy the bremsstrahlung-requirement. The use of foams would not improve significantly the volume-averaged hydrogen/boron ratio, but severely reduce average density.

than 50%, but to even less than 20% for LiBD$_2$T$_2$ and the other fuel mixes considered here at the same $\varrho_{DT,h}R_h$ and $T_h$.

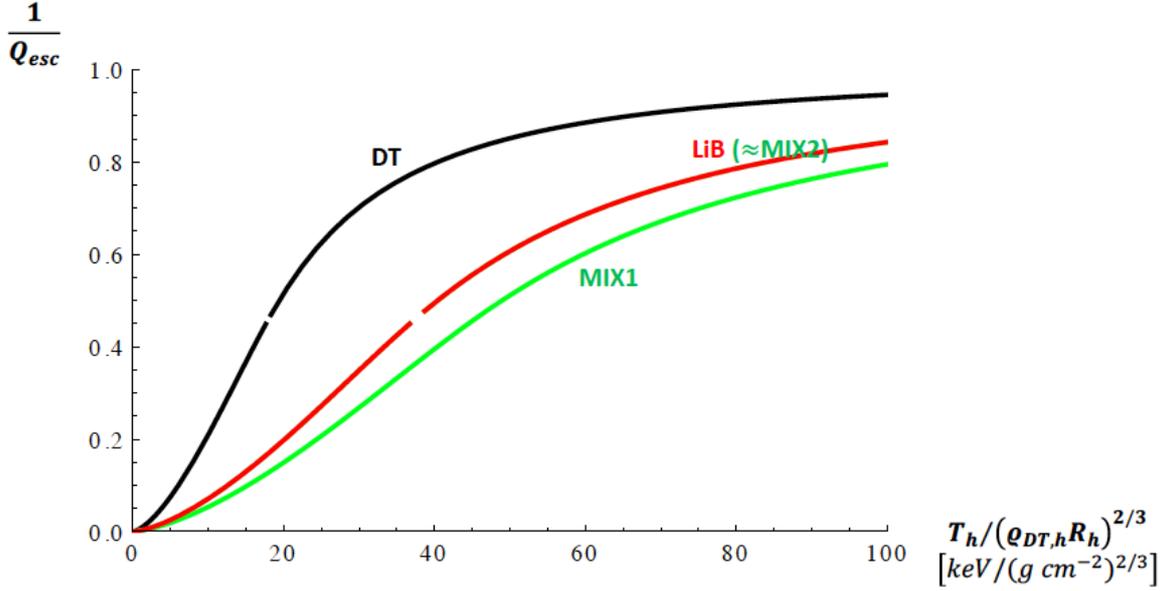

Figure 1. Escape of α-particle. $Q_{esc}^{-1}$: fraction of fusion produced charged-particle energy lost by escaping α-particles from hot-spot region for different mixed fuel compositions.

2.2. Radiation losses via bremsstrahlung

Bremsstrahlung losses are the main - generally fatal - drawback of mixed fuels. As they are caused, like fusion reactions, by binary collisions, their fractional energy loss is independent of density, but sensitive to the fuel composition. Fig.2 shows that for *DT* it is an important loss channel only as it sets the lower bound to ignition temperatures; for mixed fuels, however bremsstrahlung losses exceed fusion gain for all realistically possible solid boron compounds (and also, not shown, for lithium hydrides, which could also be a contender) except lithium-borohydride. Also, the original composition considered by RK [5] (MIX1) could not satisfy this necessary condition. However, even the existence of a window in the LiBD$_2$T$_2$ or MIX2 bremsstrahlung - to- α-particle power balance with $Q_{bre} > 1$ still implies that only 15% of the charged particle fusion energy would at best be available to cover all other loss channels.

Reabsorption of bremsstrahlung in the hot-spot region could facilitate self-heating. It would become significant when the radiation flux density at the hot-spot surface approaches the black-body limit [11], i.e., if $W_{bre} \to W_{bb} \approx \sigma_B T_h^4 \frac{S_h}{V_h} \approx \sigma_B T_h^4 \frac{3}{R_h}$ (where $\sigma_B$ is the Stefan-Boltzmann constant, and $S_h$ and $V_h$ are the surface area and the volume of the hot spot, respectively). This will evidently more easily happen for mixed fuels, and implies that $\frac{1}{Q_{bre}}$ has to remain below $\frac{W_{bb}}{W_\alpha} \sim \frac{T_h^4}{\langle \sigma v \rangle \varrho_h^2 R_h}$ (with $\langle \sigma v \rangle$ the fusion rate coefficient) for our estimates in Fig.2 to be valid. Indeed $W_{bb}/W_\alpha$ remains above 30 for all temperatures even for a strongly compressed mixed fuel (the example used for our Fig. 4) and above 40000 for the solid state density parameter set [9] of the tamped expansion scheme discussed in section 4.

Reabsorption of bremsstrahlung will thus not play a role for the mixed fuel scenarios discussed in this note.

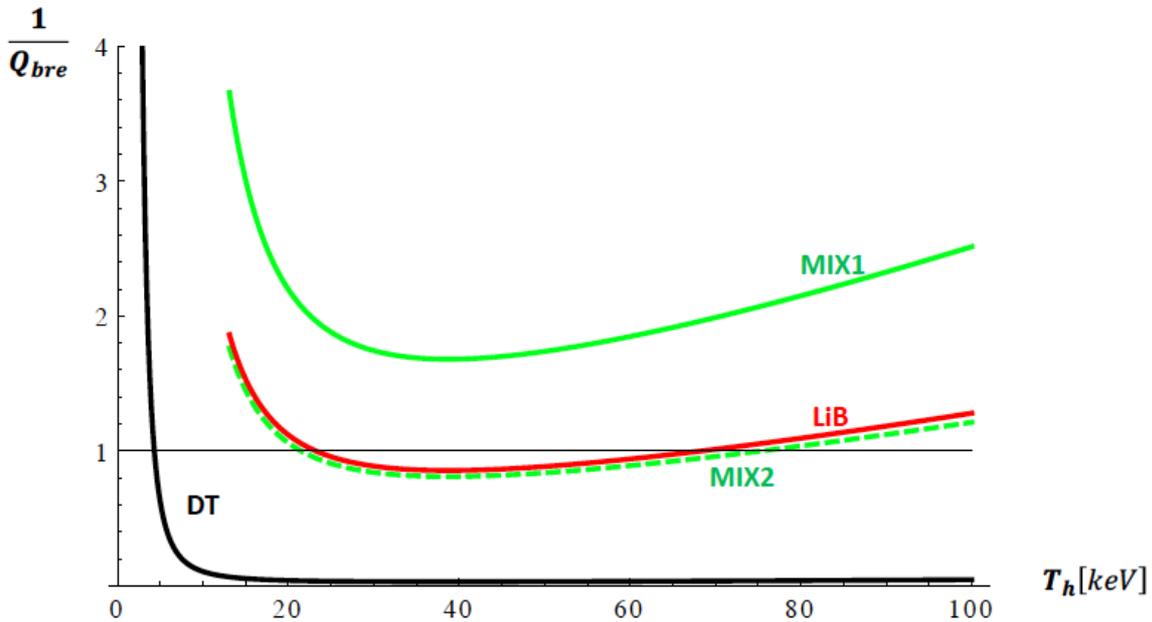

Fig.2. Bremsstrahlung losses. $Q_{bre}^{-1}$: fraction of fusion produced charged-particle energy lost by bremsstrahlung radiation from hot-spot region for different mixed fuel compositions.

2.3. Hydrodynamic expansion work

Hydrodynamic expansion losses are proportional to the ion-acoustic speed and the energy density [11]. $Q_m$ is linearly proportional to $\varrho_{DT,h} R_h$ for a given composition and temperature. A small benefit is gained by the higher inertia of mixed fuels due to the higher mass number of boron (and other constituents) [5], but this is overcompensated by the additional electrons which contribute to the pressure driving the expansion and add to the energy density of the fuel. Just accounting for this loss, even the best-performing mixed fuel (again LiBD$_2$T$_2$) would still require a $\varrho_{DT,h} R_h$ value about twice as large as pure DT (Fig.3).

2.4. Electron heat conduction

Electron heat conduction losses depend strongly on electron temperature (like $\sim T_e^{7/2}$) and weaker on the effective charge number of the plasma ($Z_{eff}$; see, e.g., Ramis [13]). Due to their diffusive nature, their resulting loss fraction is inversely proportional to $\left(\varrho_{DT,h} R_h\right)^2$. For a simple representation like given in AMtV (their equ. 4.10) a shape-factor $c_e$ of order 1 has to be specified, which we chose as 0.5 for our Fig. 4 - an optimistic value, which is at the lower end of the range proposed in [14]. At typical DT ignition temperatures and for the $\varrho_{DT,h} R_h$-values of standard laser-fusion scenarios the conduction losses are generally subdominant to expansion losses, unless the latter are suppressed, as for isobaric initial conditions or tamped expansion. For the specific cases discussed in sections 3 and 4 we include also corrections to these expressions needed when heat conduction losses start to approach the free-streaming ("Knudsen") limit (see AMtV), but these turn out to be irrelevant in the burn phase.

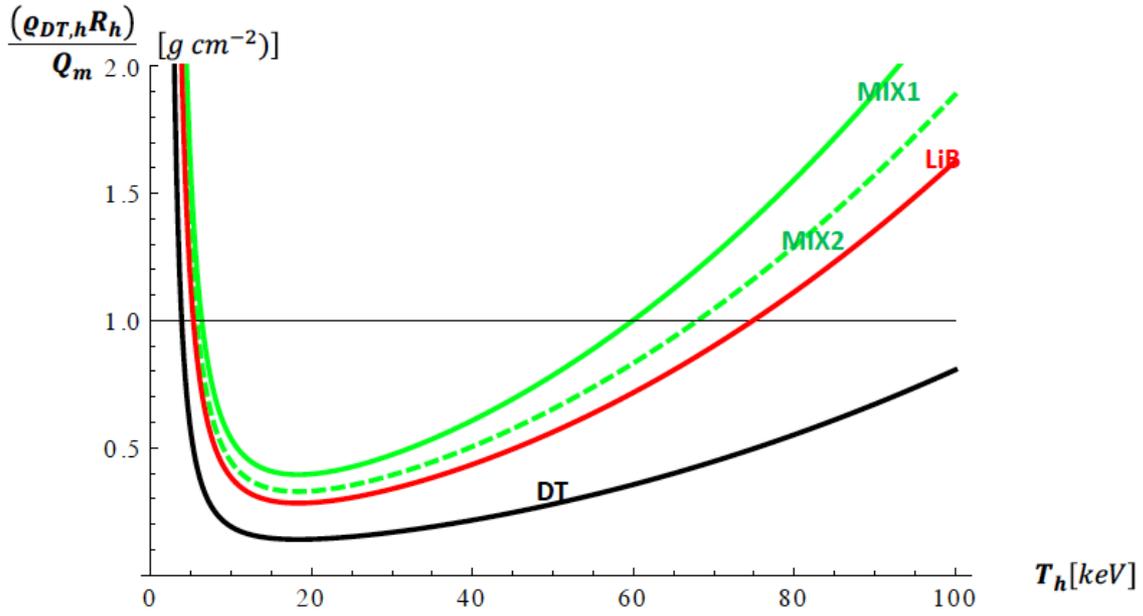

Fig.3. Expansion work. $Q_m^{-1}$: fraction of fusion produced charged-particle energy lost by thermodynamic expansion work of hot-spot against a cold background with equal density (isochoric case) for different mixed fuel compositions.

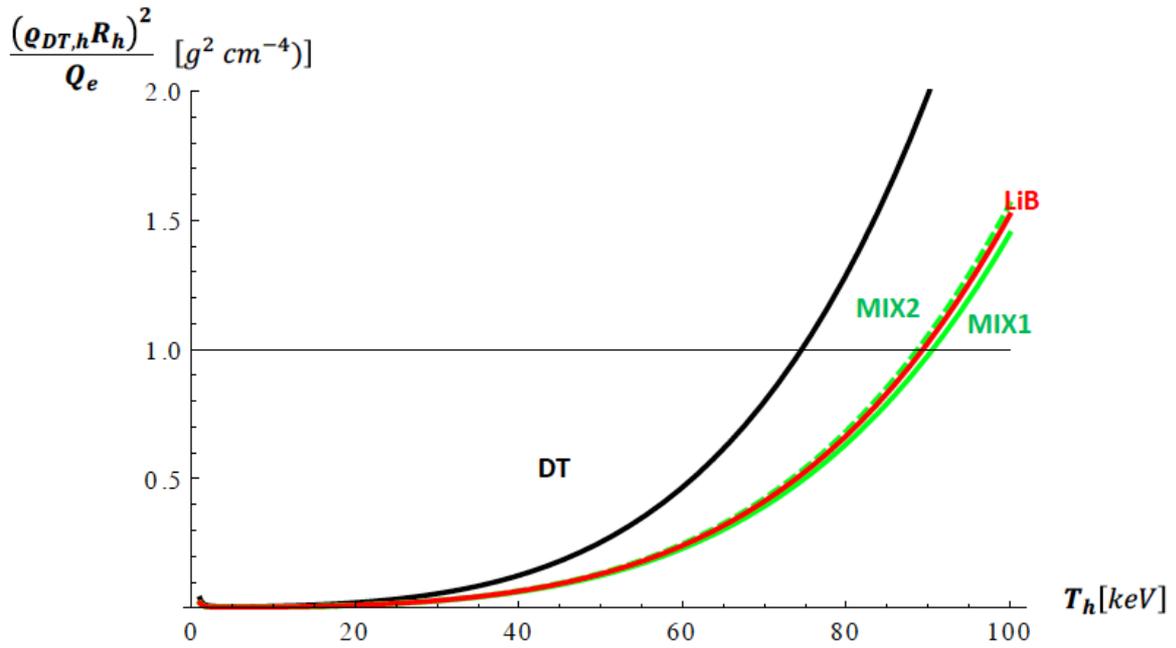

Fig.4. Electron heat conduction losses. $Q_e^{-1}$: fraction of fusion produced charged-particle energy lost by electron heat conduction from the hot-spot region for different mixed fuel compositions.

## 2.5. Comprehensive self-heating condition

The definition of self-heating given by equ. 4.17 in AMtV is equivalent to requirement of the sum of all inverses of the partial $Q$ 's to be $\leq 1$: our equ.3. For a specified $\varrho_{DT,h}R_h$ value the functions described in Figs. 1 - 4 can be summed up to yield this condition as function of $T_h$. For pure DT, and the isochoric parameter set of AMtV with $\varrho_h R_h = 0.5 g cm^{-2}$ this is done in Fig. 5. It confirms the known result, that self-heating would be possible in this case over a range of temperatures around $T_h = 12\ keV$, limited below mainly by bremsstrahlung and above by growing transparency for $\alpha$-particles, but also hydrodynamic expansion (mechanical) and conduction losses.

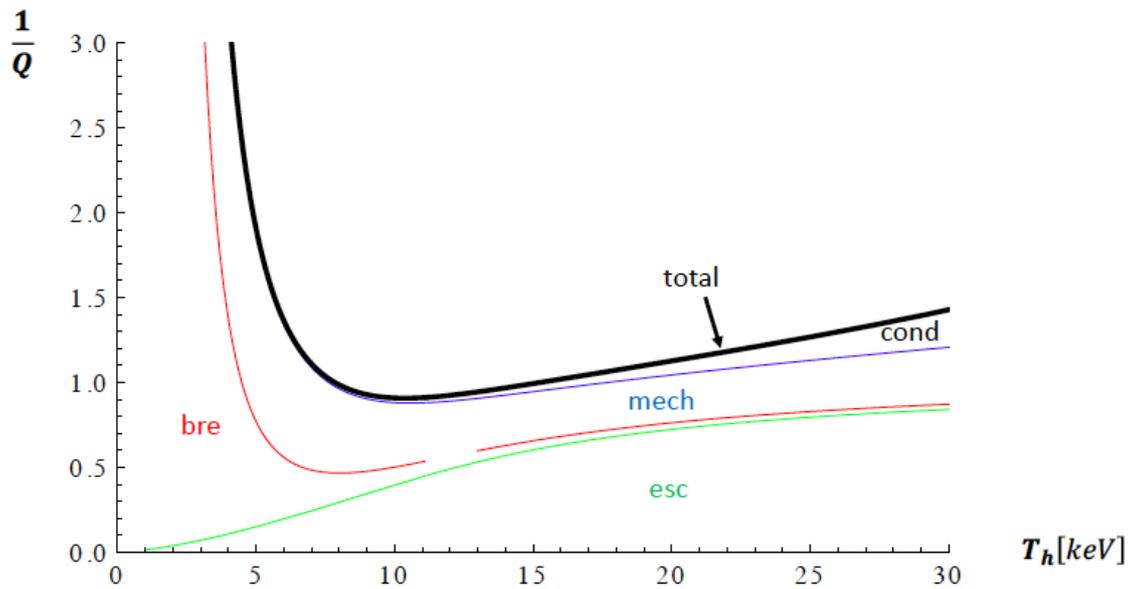

Fig.5. Losses for DT reference case. $Q^{-1}$: ratio of total energy lost out of the hot-spot to the fusion produced charged-particle energy, and contributions of the individual loss channels for pure DT fuel with an areal density of $\varrho_{DT,h}R_h = 0.5 g cm^{-2}$. The contours correspond to respective partial sums of losses in the sequence: escaping α-particles, bremsstrahlung, mechanical expansion energy, heat conduction.

For comparison with a mixed fuel, we take the only actually existing one satisfying the bremsstrahlung criterium: LiBD$_2$T$_2$, and show the contributions to $1/Q_\alpha$ *at the same value of* $\varrho_{DT,h}R_h$ *as for the pure DT reference case* in Fig.6. Even under this condition - which implies several hundredfold compression - self sustained burn is not possible, with the optimum achievable $Q_\alpha$ always below 0.5. This is primarily due to the dramatically enhanced bremsstrahlung-losses, which also shift the ignition requirements to much higher electron temperatures, annihilating thereby the advantage mixed fuels have regarding $\alpha$-particle stopping! Mechanical expansion losses are also enhanced over the pure *DT* case, but would alone not preclude self-heating at these large $\varrho_{DT,h}R_h$ - values.

Note, that this comparison set is made at equal $\varrho_{DT,h}R_h$ assumption, but equal $\varrho_{DT,h}$ at given temperature would imply in addition a more than double plasma pressure and hot spot energy for the mixed fuel case.

Particularly the dominant contribution by bremsstrahlung is a consequence of the constraint that the starting material should be a solid at room temperatures. This is in turn, however, one of the advantages quoted for the mixed fuel concept. The second was the possibility to use DT reactions as a trigger to arrive ultimately at practically neutron-free pB fusion. However even estimates by RK [5] showed that self-sustained burn of pB would require ion temperatures in the 200 keV range, and about a factor 1000 gain in the DT stage, and any reference to contributions from pB reactions has been dropped in the more recent scientific notes by MARVEL Fusion authors.

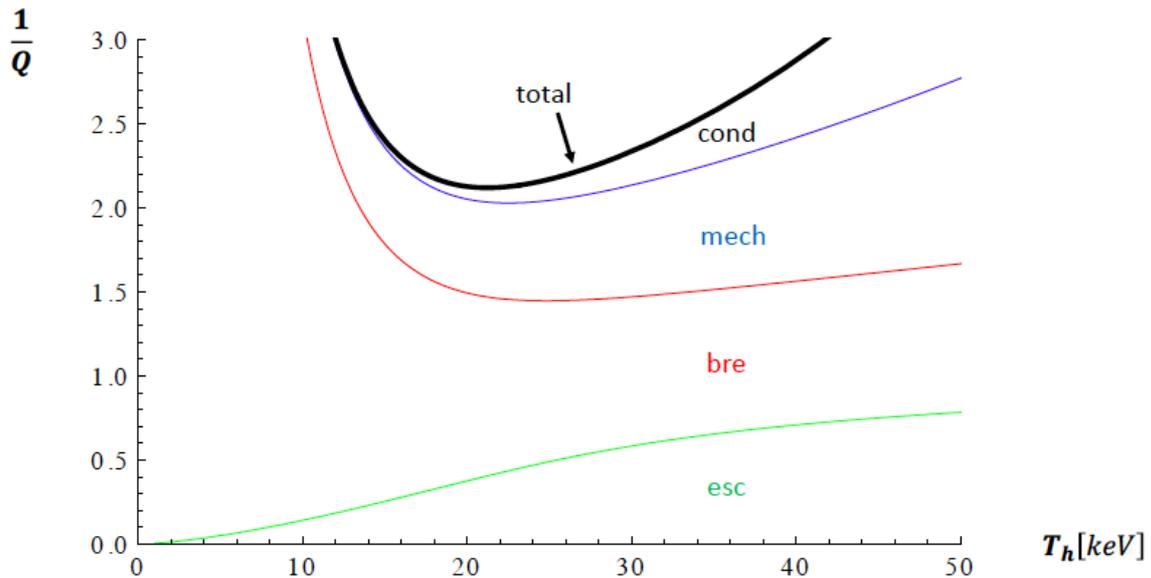

Fig.6. Losses for compressed "heavy" lithium-borohydride. $Q^{-1}$: ratio of total energy lost out of the hot-spot to the fusion produced charged-particle energy, and contributions of the individual loss channels for „heavy" lithium- borohydride (LiBD$_2$T$_2$) with an areal density of $\varrho_{DT,h} R_h = 0.5 g cm^{-2}$. The contours correspond to respective partial sums of losses in the sequence: escaping α-particles, bremsstrahlung, mechanical expansion energy, heat conduction.

Due to the much-enhanced bremsstrahlung losses and the resulting unfavorable shift in optimum reaction temperatures to much higher values, mixed fuels dramatically enhance the difficulties for reaching self-heating, ruling it practically out for all known solid hydrogen compounds including the only one for which fusion could marginally beat bremsstrahlung: LiBD$_2$T$_2$. The need for much higher $\varrho_{DT,h} R_h$ would dwarf the advantage of an up to 40% higher natural hydrogen mass density than solid hydrogen, particularly considering the higher pressures and energy densities associated with them.

### 3. Analytic Estimates for Self-Heating in Uncompressed Fuels

Besides a possible access to aneutronic fusion, and an improved coupling of high-power laser energy the possibility of achieving fusion self-heating and high energy gain without the need of precompression was a central driver for the interest in mixed fuels by RK. Arguments given for this were the reduced escape of $\alpha$ particles and a reduced rate of fuel disintegration due to the higher inertia of boron.

To check these claims against our results of the preceding section, we show in Fig.7 the loss channels for the optimum mixed fuel LiBD$_2$T$_2$ at $\varrho_{DT} = 0.3 g/cm^3$ (corresponding to its natural hydrogen content in solid state), assuming a fuel radius of 2.5 mm to give the same $\varrho_{DT,h} R_h = 0.075 g/cm^2$ as the optimum value with the mixture b3 (our MIX1) in Fig.3 of ref. [5]. The results clearly preclude significant fusion self-heating for this parameter range, with a predicted maximum $Q_\alpha$ of 0.1.

The presence of boron and lithium indeed reduces $\alpha$-particle and heat conduction losses, but these effects are overcompensated by the lower areal density and - in particular - the shift of the optimum operation point to much higher values of temperature enforced by the nearly 30fold increase of bremsstrahlung. In fact, because of this synergetic effect bremsstrahlung and $\alpha$ escape alone ensure the impossibility of significant self-heating for any practicable fuel dimension.

Hydrodynamic expansion is actually also *enhanced*, as the pressure of electrons largely compensates the benefits of enhanced mass density in the expansion velocity, and adds to the heat capacity. The mechanic expansion work alone would suffice to keep $Q_\alpha < 0.2$. Compared with Fig. 6 these results confirm that strong fuel compression would be needed also in the mixed fuel case to reduce $\alpha$ particle losses, hydrodynamic expansion work and heat conduction to a level allowing for significant fusion self-heating, although even this might be futile as at best 15% of fusion $\alpha$ particle energy would be available to cover these non-bremsstrahlung losses.

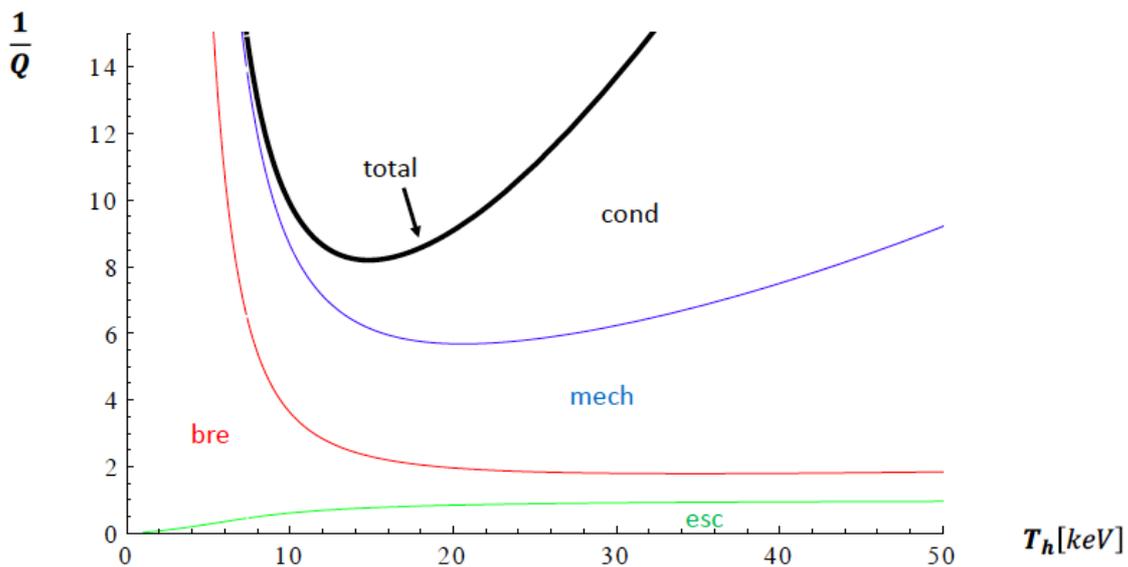

Fig.7. Losses for solid-density „heavy" lithium-borohydride. $Q_\alpha^{-1}$: ratio of total energy lost out of the hot-spot to the fusion produced charged-particle energy, and contributions of the individual loss channels for „heavy" lithium- borohydride (LiBD$_2$T$_2$) with an areal density of $\varrho_{DT,h} R_h = 0.05 g cm^{-2}$. The contours correspond to respective partial sums of losses in the sequence: escaping α-particles, bremsstrahlung, mechanical expansion energy, heat conduction.

The need for compression arises from the requirement to attain a sufficient $\varrho_{DT,h} R_h$ value with realistic values for the initial hot-spot (and hence laser-) energy $E_i$, which scales as $\varrho_{DT,h}^{-2}$. In fact, for pure *DT*, AMtV give an expression for the needed hot-spot ignition energy (equ. 4.24) which applied to the natural density of *DT*-ice ($\varrho_{DT} \approx 0.22 \ gcm^{-3}$) would yield 20 GJ.

RK, making simple analytic estimates of hydrodynamic and radiation losses [5], came to dramatically more optimistic conclusions about the needed hot-spot energy and the achievable $Q$ (equivalent to our $Q_\alpha$): " The natural proton density in the chemical compounds is about $\varrho_p \approx 10^3 kgm^{-3}$ implying $E_i \approx 1.0 \ MJ$ and $P_\tau \approx 10^{19} \ W$ for a $Q \approx 1.0$ and $kT_i \approx 30 \ keV$ ignition according to *(their)* Fig. 4."

On the way to this extreme result three unjustified assumptions were made by RK in [5], which lead to a more than three orders of magnitude difference between these predictions.

(1) In estimating mechanical expansion losses and in ignoring heat conduction, RK neglected electron temperature, which would contribute to the expansion speed and the energy content and added heat losses
(2) They did not consider in their estimate quantitatively the escape of $\alpha$ particles
(3) They estimate loss channels (mechanical expansion losses and bremsstrahlung) separately. In fact, the loss channels (plus heat conduction) sum up, and act, in particular in a synergetic way.

*It is in fact this negative synergy among the loss channels which invalidates the results of RK's model and any claims for advantages of mixed fuels: the dominance of bremsstrahlung over fusion power at low temperatures forces operation to higher temperatures, where the plasma gets more transparent for $\alpha$ particles and hydrodynamic and heat conduction losses increase strongly (compare our Figs. 5 and 6).*

In their most favorable estimate for the Laser energy requirement without compression they moreover postulated an unrealistically high natural hydrogen density in the mixed fuel of $\varrho_p \approx 10^3 kgm^{-3} = 1 \ gcm^{-3}$, which is more than 10times that of hydrogen ice ($\varrho_p = 0.086 \ gcm^{-3}$. Boron hydrides can indeed exceed the value for hydrogen ice, but only by a much smaller factor (LiBH4 has $\varrho_p \approx 0.12 \ gcm^{-3}$). In the initial version of their note [5] RK used a composition $n_D = n_T = n_p = n_B$ (labelled MIX1 in our Figs.1 -4) for which actually bremsstrahlung alone would have precluded thermonuclear burn (see our Fig.3). They have corrected this composition now to $0.6 \ n_D = 0.6 \ n_T = 0.6 \ n_p = n_B \ (MIX2)$, which marginally satisfies $Q_{bre} > 1$, but appears completely unrealistic in its favorable gravimetric hydrogen density as solid.

The common driving physics assumption in this estimate in [5] is in fact the neglect of (or in case of the bremsstrahlung, the use of a much reduced) electron temperature and its consequences. This would be justified in the regime of several hundred keV considered by Putvinski for pure *pB* burn[iv] [12], but not in the few tens of keV regime of dominating DT fusion, where the stoppage of $\alpha$ particles happens predominantly on electrons. At electron

---

[iv] without the constraint to solid materials, Putvinski can propose a much more hydrogen-rich fuel composition which might open at about $T_i \approx 300 \ keV$ even a window for B-p fusion power to exceed bremsstrahlung.

temperatures $T_e \leq 20 keV$ more than 75% of the $\alpha$- particle energy goes directly into heating the electrons [15], and will be available for net ion-heating and the drive of fusion burn only if $T_e$ becomes $> T_i$ . Heat conduction and bremsstrahlung will tend to reduce electron temperatures, but the scarcity of channels for preferential ion heating (a small fraction of the fusion-power and primarily shock waves) and collisionality will keep both temperatures close together.

This argument of preferred electron heating refers not only to the burn phase where plasma heating is predominantly due to fusion-$\alpha$'s, but also to the laser heating stage, if the thermal plasma is heated by fast particles. In fact, in later papers of RK ([6], [9]) these assumptions - specifically the neglect of electron temperature - have been dropped. As now also RK agree [9], recognized models [8] predict that: "without energy feedback[v] and enhanced fuel confinement $E_i > 1GJ$ would be needed for $Q_F \approx 1$."

### 4. A concept of tamped expansion

A major loss channel for an uncompressed fuel is mechanical expansion work. This could be reduced by tamping the expansion by cladding the fuel with material of high mass density. This principle is well known and in everyday use in explosions, and elements of it are adopted also in "fast-ignitor" proposals (the shielding of the ignition laser path by a high mass-density protective cone, see, e.g. AMtV). It is based on the idea that at the interface of materials with vastly different mass density, pressure waves from the low-density side will be reflected. Essentially the disintegration will be stopped, and no expansion work by the fluid would have to be done. Evidently ignition would become easier - the analytic criterium reduces to the isobaric case in AMtV's Fig.4.2, which gives a required $(\varrho_{DT,h} R_h)$ by a factor of 2 (see AMtV's formulae 4.21 and 4.23) lower than the isochoric one. This is the basis of the latest fusion target proposal by RK [9], in which the fuel is cladded in a gold layer.

This cladding in the configuration presented in [9] is limited, however, only to the mantle of a cylinder, leaving open top and bottom, driven by the need of laser access. The height of the cylinder $(L_h)$ is assumed to be equal to the fuel radius, which would actually lead only to a reduction of expansion work by about 1/3 compared to a spherical arrangement based on a crude estimate of the uncladded surface to the volume. But even neglecting expansion losses completely, the benefits would be limited by the persistence of the other loss channels: neither $\alpha$- particles, nor bremsstrahlung radiation nor electron thermal heat flux will be reflected on the high-mass density surface.

But even ignoring the persistent expansion losses, the required hot-spot energy for self-heating according to AMtV (their equ.4.22) for the relevant isobaric case would still be 250 MJ for a pure $DT$ mix with the "uncompressed" density postulated by RK of $0.5 g/cm^3$. Inserting instead the $DT$ density $0.3\ g/cm^3$ corresponding to lithium borohydride as most suitable hydrogen carrier described in the literature, this energy would increase to 700 MJ, and this estimate would still not account for the about 30-fold increase in bremsstrahlung's-losses for LiBD$_2$T$_2$.

---

[v] significant energy feedback arises only once $Q_\alpha$ approaches 1

For a more comprehensive picture taking account of the mixed-fuel situation, we apply the formalism of section 3 to the case described in most detail by RK in [9] (corresponding to their Figs. 4, 5), using their fuel mix ($0.6 n_D = 0.6 n_T = 0.6 n_p = n_B$) with $\varrho_{DT,h} R_h = 0.05\ g/cm^2$ and omitting any expansion losses. Our Fig. 8 shows, that bremsstrahlung's losses alone will keep $Q_\alpha < 1$ for temperatures below 20 keV, at which temperature the fraction of escaping $\alpha$-particle is already 80%, and heat conduction losses are 4 times larger than the total fusion power in charged particles.

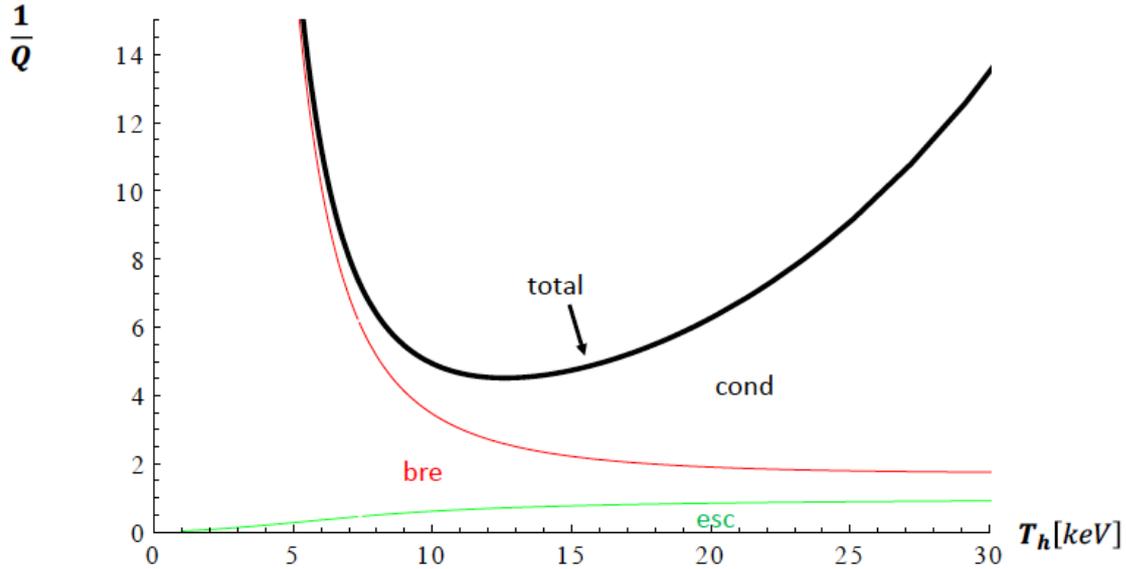

Fig.8. Losses for solid-density MIX1 and tamped expansion. $Q^{-1}$: ratio of total energy lost out of the hot-spot to the fusion produced charged-particle energy, and fractional contributions of the individual loss channels for a scheme with perfectly tamped expansion, for the mixed fuels described by RK in ref. [9] ($0.6 n_D = 0.6 n_T = 0.6 n_p = n_B$) and the areal density in their figs.4,5: $\varrho_{DT,h} R_h = 0.05\ gcm^{-2}$. The contours correspond to respective partial sums of losses in the sequence: escaping $\alpha$-particles, bremsstrahlung and heat conduction.

We examine also the effect of an increase in the hot spot disk radius by a factor of 4 within our model (Fig. 9.). This would indeed reduce dramatically the losses due to heat conduction and significantly those due to $\alpha$ − escape - but the now dominant role of bremsstrahlung would still keep $Q_\alpha$ near 0.5, illustrating the diminishing return from further size increase. (Already this disk size, assuming $L_h = R_h$ would require an initial hot-spot energy of about 500 MJ to be provided by Laser energy deposition.)

It is known from full 1-d simulations of Laser plasma interaction that in a truly isobaric situation, ignition is possible for $\varrho_h R_h$ also below the value given by equ. 4.21 of AMtV (i.e., $Q_\alpha = 1$) if the tamping material surrounding the hot spot is cold $DT$ fuel at much higher density (i.e., corresponding to equal pressure, achieved by strong compression). In this case the heat conduction flux and escaping $\alpha$- particles out of the hot spot are not actually lost, but can heat this surrounding fuel. This beneficial effect, however, will not arise for RK's scheme of tamped explosion, where $\alpha$ - particles, heat conduction and bremsstrahlung will deposit energy not in fuel, but in a cladding of gold.

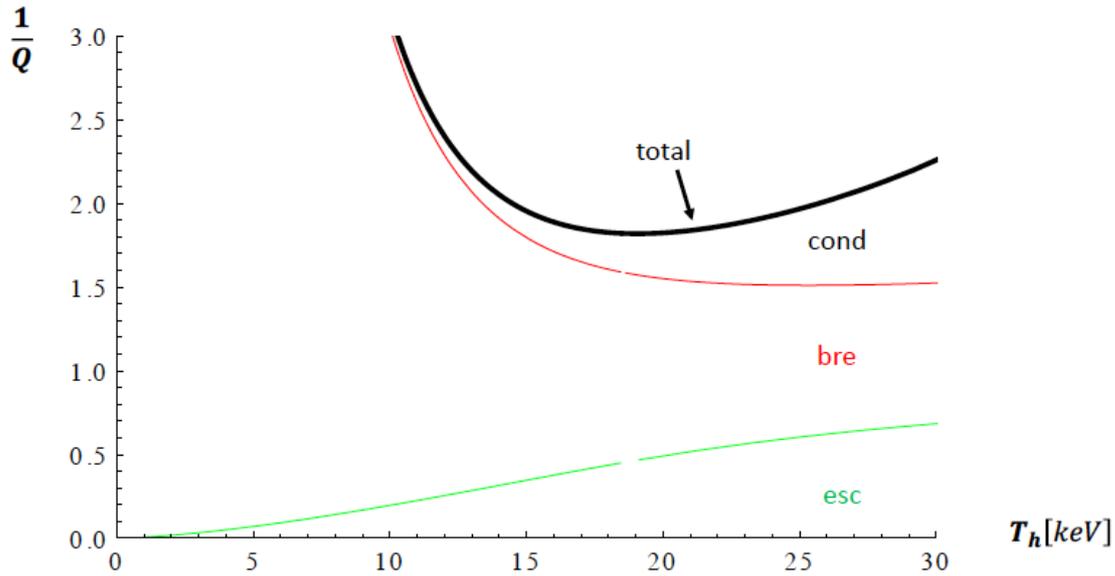

Fig.9. Effect of target size increase for tamped expansion of an uncompressed target. $Q^{-1}$: ratio of total energy lost out of the hot-spot to the fusion produced charged-particle energy, and fractional contributions of the individual loss channels for a scheme with perfectly tamped expansion, for the mixed fuels described by RK in ref. [9] ($0.6n_D = 0.6n_T = 0.6n_p = n_B$) and but a fourfold increased areal density: $\varrho_{DT,h}R_h = 0.2 g cm^{-2}$. The contours correspond to respective partial sums of losses in the sequence: escaping α-particles, bremsstrahlung and heat conduction.

These pessimistic results are in apparent contradiction to statements of RK [9], this time based on the community code MULTI [13], which is quoted to predict fusion target yields of up to 4 for deposited laser energies of 45 MJ. For presenting summary results of their simulations the authors use a quantity $Q_F$ as the ratio of the final energy in the reactor to the deposited initial one. - This definition would give $Q_F = 1$ even in the absence of any fusion reactions, as long as the heat flux and the radiation from the originally hot fuel remain trapped in the reactor. From equ. 2-4 of ref. [9] it is however clear, that this is not intended, and only the *additional energy* in the target due to fusion reactions is to be counted. From these formulae and the explicit statement that the quantity $\epsilon_f$ used in them refers to "the elementary fusion energies of all fuels involved *without neutrons*" it is also evident that $Q_F$ measures the energy deposited in the target rather than in the reactor.

The definition of $Q_F$ requires in general the output of a dynamic simulation code for evaluation and is not directly comparable to our $Q_\alpha$. Sufficiently below $Q_\alpha = 1$, however, an approximate relation exists between the two quantities - $Q_\alpha \approx \frac{Q_F}{1+Q_{esc}Q_F}$ - in which the fraction of α-particle energy escaping the fuel ($Q_{esc}$) appears explicitly as they are considered confined in the target but not in the fuel.

MULTI is a well established and well tested code, but used here in an unconventional set-up by non-authors. The mixed nature of the fuel - which cannot be directly handled by the used version of the code - is mocked-up by alternate layers of DT and pB with a global (average) mix of $n_B = 0.6n_p = 0.6n_D = 0.6n_T$. More details of these simulations are given for one particular case, corresponding to the parameters used also for our fig. 8. Directly shown graphical output, and statements in text and figure captions in ref. [9] are, however, in mutual contradiction regarding critical quantities. $Q_F$ as defined by RK can be directly read-off from MULTI-profiles of energies before (Fig. 4 in ref. [9]) and towards the end of burn (Fig.5 in ref. [9]). This procedure yields (based on a volume integration of the graphically shown energy densities and including the energies in the gold-cladding [vi]) a gain of a few tens of percent, in better agreement with our estimate of 0.2 than the value of $Q_F = 2$, claimed in the figure caption of ref. [9]. Our estimates of $Q_F$ and $Q_\alpha$ as far below any onset of significant self-heating [vii] is also consistent with the cooling of the fuel and the observation that most of the energy (initial + fusion energy) after 1 ns is in the gold layer.

An indication for the origin of this discrepancy might follow from further output parameters of the calculations shown in Fig. 5 of ref. [9]. The fusion power density $R_f$ given there is 20x larger than the $\alpha$- particle power following from the shown pressures and ion temperatures. A factor of 5 could be due to an inclusion of the neutron energy in this definition, but only a further factor of 2 can be justified on account of the fact that in a layered simulation the fusion power should be corrected for the effective reduction of fuel volume. Moreover, the same enhancement factor should then also be applied to the thermal capacity and the quoted initial hot spot energy, so that this correction, consistently applied, would not lead to an enhancement of $Q_F$.

Due to a more subtle effect the procedure used to account for complex fuel mixtures in the MULTI-simulations will overestimate fusion power by a further factor of 1.2 and the ratio of bremsstrahlung's losses to fusion power production by a factor of the order 1.4. It arises from the fact that the alternating fuel layers are - according to fig.4 of ref. [9] - evidently not in initial pressure equilibrium, due to the spatially uniform electron temperature but higher electron density in the pB layers. This will immediately equilibrate, but lead in this process to a compression of the DT layer by 1.2, and an expansion of the pB layer by an equal factor. As fusion happens in the DT, but bremsstrahlung's-losses in the pB layer, this will artificially enhance fusion by a factor of 1.2 and reduce bremsstrahlung by the same factor, as both processes go quadratically with density and but only linearly with volume. -

We therefore believe that these MULTI simulations do not disprove our estimates: tamping - even if it could be achieved in all directions without impeding laser access to the fuel - would not eliminate the need for strong precompression of it. Using a mixed fuel, even of optimum composition, would further increase these difficulties due to the at least 30fold increase in

---

[vi] for this we assume that not only bremsstrahlung and heat flux, but also $\alpha$ particles escaping the fuel are absorbed in the gold-cladding. This is well justified as the energy stopping length for 3.5 MeV $\alpha$'s in gold (including properly the contribution from bound electrons) is below 10 $\mu m$ according to (https://physics.nist.gov/PhysRefData/Star/Text/ASTAR.html)

[vii] the simple way for the proponents to prove significant self-heating would have been to show results of a comparison run with fusion reactions switched off.

bremsstrahlung losses, by far outweighing any gain from a somewhat higher initial hydrogen density and higher collisionality and inertia of the fuel.

## 5. Marvel Fusion's comments to our criticism

In their latest note [9], RK address some of the critical remarks we made in previous comments. They specifically question the following statements by us, which we insist, however, to be correct and relevant:

> 1. *Compression is always needed and necessary*. - RK claim that this is mainly due to the need to stop $\alpha$- particles, which mixed fuels indeed do more effectively, owing to the higher number of electrons. This is, however, only *one* criterium driving the need for high $\varrho_h R_h$ and hence compression. The other - equally necessary requirements come from delaying disintegration (= expansion work) of the fuel and from reducing electron heat conduction. While the former can be affected by tamping - subject to compatibility with Laser access -electron heat conduction will persist. It gives generally a less stringent criterium than the expansion work (see difference between equ. 4.21 and 4.23 in AMtV), but is still far from tolerable without compression.

RK also argue that textbook knowledge, based on pure DT fuel are not applicable to mixed fuels, as the latter ignite at much higher temperatures. - *But exactly as they can ignite only at higher fuel temperatures (and at actually also higher $\varrho_h R_h$), mixed fuels would require prohibitively large ignition (i.e., Laser-) energies.*

> 2. *Ultra-intense laser pulses bring no benefit* (unless accompanied by strong precompression of the plasma). - RK's analyses of thermonuclear burn start from the assumption that the preheating takes place on a time much shorter than the confinement time $\Delta \tau$, which for their parameters (without precompression) is on a nanosecond scale. In this case there is no benefit in reducing the Laser pulse length much below 100 picoseconds (unless the laser-fuel coupling scheme requires this, but then at the cost of higher power and hence more advances technology). The situation is different for a strongly precompressed fuel as in a fast ignitor regime, where the relevant confinement time is in the 10-100 ps range, and a commensurately shorter laser pulse is needed.

> 3. *Rocket efficiency speaks against ultra-fast laser pulses*. - In fact, this is a consequence of momentum conservation, and not violated by any Laser-matter interaction scheme. (Impact fusion - in its different forms - carries momentum from the cluster or the EMF accelerator to the target, and the recoil is sustained by the accelerator). The laser light carries negligible net momentum from the source to the target: the momentum is created only in the laser-plasma interaction, either through thermal expansion or through the ponderomotive force [16]. In both cases the momentum available for compression is equal to the one of the mass accelerated in the opposite direction, and hence profits if the latter is large. (This is actually also illustrated by the calculations of RK with the MULTI code, which show compression only when the high mass-density gold cladding is ablated by heat conduction and bremsstrahlung absorption *on a ns time scale*, much longer than the femtosecond laser pulse time.)

In RK the authors furthermore discredit our criticism by pointing to our lacking experience in the field of ICF. This is actually wrong, as K. Lackner wrote one of the first hydrodynamic Laser fusion codes for non-military applications (time dependent Lagrangian, spherically symmetric, including alpha-particle heating and electron heat conduction), gave the first invited talk on the subject of inertial fusion at the DPG, and presented there also his results on the need for strong compression [17]. Moreover, our criticism has been confined to issues of general plasma physics and nuclear reactions, where a large common ground exists between ICF and MCF.

*In particular, however, the relevance of our criticism has been implicitly also recognized by RK, as - following respective comments by us - they have retracted several earlier claims in later notes, or even submitted revised versions of previous ones. In particular those points were:*

- in [2] RK presented „a non-thermal laser-driven mixed fuels nuclear fusion reactor concept". -  We [18] pointed to the fact, that these beam-target reactions would not play a role, unless the $(\varrho R)$ product of the surrounding hot plasma were sufficiently high. - *RK produced a revised version of this publication in which reference to the non-thermal reactions was eliminated from the title [3]. - In later publications RK [9] even stressed the fact that fusion reactions have to be thermal and neglected any direct contributions due to laser produced fast particles.*

- in [5] RK neglected electron contributions in the estimates of expansion energy losses and assumed $T_e = 0.5\, T_i$ in the bremsstrahlung losses in their analytic estimates of fusion yield. - We pointed to the fact that *DT*-fusion $\alpha$-particles predominantly heat electrons, so that this assumption was not justified [19,20].- *RK issued a revised version of this note, where $T_e$ was raised to $T_e = 0.75\, T_i$ [6]. In later estimates RK assumed $T_e \approx T_i$, as confirmed also by their own simulations with the MULTI-code [9].*

- in [2] RK invoked self-generated magnetic fields to confine the laser-produced plasma. - We [18] pointed to the fact, that the well-established virial-theorem excludes confinement by purely internally generated magnetic fields. - *RK issued shortly after a revised version of this note, in which the corresponding section was dropped [3].*

- in [21] RK suggested that radiation losses could be largely reduced, by confining the plasma in a cavity with a radiation temperature equal to the electron temperature. -  We [20] pointed to the fact that in the thermonuclear regime this would imply a latent energy in the radiation field largely surpassing the thermal one of the plasma. - *RK issued shortly after a revised version of this note, in which this suggestion was dropped [6] and never raised it again in later notes.*

- aneutronic fusion plays a pivotal role in the earlier presentation by MARVEL scientists [7]. - From our very first note we stressed that (pB) fusion reactions would require prohibitive hot spot energies and therefore also imply (for a predominantly aneutronic energy production) an unmanageable single-event explosive energy release [18]. We [19] specifically pointed to the fact that RK's own estimates (their Fig. 4 in ref. [5]) predicted a 1000-times larger ignition energy for pB than for DT. - *A revised version of this particular note was subsequently submitted by RK [6], in which these calculations concerning pB were omitted, and pB fusion power contributions have been dropped in all subsequent notes of RK addressed to the scientific community.*

RK also stated in [10] that we [19] wrongly implied that their calculations reported in [5] assumed also the neutron energy to contribute to plasma self-heating. - *This is correct,* and our mistake was triggered by their use of the symbol $\epsilon_f^{DT}$ which was defined in their earlier publication [2], table 1, numerically as 18.35 MeV (i.e., the total fusion energy including the part in fast neutrons) but used in [5] only for the part carried by the $\alpha$-particles (as clarified only in a later version of their note [6]). In fact, the factor of about 5 discrepancy in the $(\varrho R)$-requirements between their formula and the textbook expressions of AMtV is due to their neglect of the electron temperature effects (and hence also bremsstrahlung and electron heat conduction-losses) as we discussed in section 3. Dropping this invalid assumption, also RK now accept the conventional results and quote $E_i > 1\ GJ$ [viii] as needed for uncompressed pure DT targets in the absence of tamping of fuel disintegration.

## 6. Conclusions

The use of mixed fuels in inertial confinement schemes severely raises the energy requirements for achieving high gain or a burning plasma state. This is primarily because of the strongly enhanced bremsstrahlung-losses, but also due to the energy needed to heat the additional electrons which raise the needed hot-spot energy and increase the hydrodynamic expansion losses. For all compounds solid at room temperature only "heavy" lithium borohydride (LiBH4) appears capable to beat bremsstrahlung losses by fusion $\alpha$ - particle heating in the optically thin regime. Even for this compound, however, the negative synergy between the different loss channels would preclude self-heating at any hot spot areal density $\varrho_h R_h$ and temperature $T_h$ for practicable pre-heating energies: *the dominance of bremsstrahlung over fusion power at low temperatures forces operation to higher temperatures, where, however, the plasma gets more transparent for $\alpha$ particles and hydrodynamic and heat conduction losses increase strongly (compare our Figs. 5 and 6).*

Uncompressed, even pure DT would require GJ laser energies for achieving ignition [8] as has now been recognized also by RK [9]. Mixed fuels, due to the enhanced losses from bremsstrahlung and mechanical work of fuel expansion would even severely raise the difficulties for reaching self-heating. The need for much higher $\varrho_{DT,h} R_h$ and $T_h$ would dwarf the potential advantages of an about 40% higher starting *DT* density than solid hydrogen.

To overcome the $\varrho_h R_h$ - constraint posed by pellet disintegration (i.e., the energy losses due to expansion work), RK have recently proposed the concept of a tamping the fuel expansion by cladding the fuel by a layer of gold. In the described geometry of a disk with axial diameter

---

[viii] we use the initial hot-spot energy $E_i$ as a proxy for the required laser energy: it would be equal to it, if laser energy could be converted into thermal fuel energy with an efficiency 1.

half the radial one, this would actually reduce expansion losses only by a modest fraction. However, even if axial tamping were perfect (which appears difficult to reconcile with the need for laser access to the fuel) this would still only avoid losses due to expansion work, and not reduce the remaining three loss channels: $\alpha$- particle escape, bremsstrahlung, and electron heat conduction, which would lead to similar constraints on $\varrho_h R_h$. In contrast to fast ignition concepts (which also profit from tamping of the hot spot plasma, but by high density cold fuel surrounding it) where heat conduction and $\alpha$- particle escape also have a beneficial effect of heating additional denser fuel, they are net losses in this case as they go into a non-reacting cladding material (gold).

The use of mixed fuels therefore does not eliminate the need for strong precompression of the fuel: in fact, it renders achieving burning plasma conditions much more difficult, if not impossible.

The original goal of mixed fuels - access to a state of pure pB burning - has been completely dropped in the later notes addressed to a scientific community. The latest reference to pure pB fusion in a paper RK is in ref. [5], where estimates in their Fig. 4, (case b5) confirmed the textbook knowledge that the start-up energy requirements would exceed those for DT by a further factor of 1000.

[1] Zylstra, A.B., Hurricane, O.A., Callahan, D.*A. et al.*, Burning plasma achieved in inertial fusion. *Nature* **601**, 542–548 (2022). https://doi.org/10.1038/s41586-021-04281-w
[2] Ruhl, H. and Korn, G., **"A non-thermal laser-driven mixed fuel nuclear fusion reactor concept"**, **[v1]** Mon, 7 Feb 2022, https://arxiv.org/abs/2202.03170v1
[3] Ruhl, H. and Korn, G., **"A laser-driven mixed fuel nuclear fusion reactor concept"**, **[v5]** Fri, 22 Apr 2022 , https://arxiv.org/abs/2202.03170
[4] Ruhl, H. and Korn, G., **"High current ionic flows via ultra-fast lasers for fusion applications"**, [v1] Sun, 25 Dec 2022, https://doi.org/10.48550/arXiv.2212.12941
[5] Ruhl, H. and Korn, G., **"Low Q nuclear fusion in a volume heated mixed fuel reactor"**, **[v1]** Mon, 13 Feb 2023, https://arxiv.org/abs/2302.06562v1
[6] Ruhl, H. and Korn, G.**, "Volume ignition of mixed fuel", [v7]** Sun, 3 Sep 2023, https://doi.org/10.48550/arXiv.2302.06562
[7] Korn, G.; The future begins with pB11 (2021), https://marvel-fusion.medium.com/the-future-begins-with-pb11-fusion-fuel-1a2a015818e8
[8] Abu-Shawareb H. et al., Lawson criterion for ignition exceeded in an inertial fusion experiment, Phys. Rev. Lett. 129 (2022 075001), https://doi.org/10.1103/PhysRevLett.129.075001
[9] Ruhl, H. and Korn, G., **"Numerical validation of a volume heated mixed fuel concept"**, **[v4]** Sun, 2 Jul 2023, https://doi.org/10.48550/arXiv.2306.03731
[10] Ruhl, H. and Korn, G., **"Comments on the comments by Lackner et al. on the series of papers about "A novel direct drive ultra-fast heating concept for ICF""**, **[v1]** Wed, 20 Sep 2023, https://arxiv.org/abs/2309.11493v1
[11] Atzeni,S, J. Meyer-ter-Vehn, The Physics of Inertial Fusion, Oxford Science Publications (2004)
[12] Putvinski S.V. et al., Fusion reactivity of the pB11 plasma revisted, Nucl. Fusion 59 (2019) 076018, https://doi.org/10.1088/1741-4326/ab1a60
[13] Ramis R., Meyer-ter-Vehn J., MULTI-IFE – A one-dimensional computer code for Inertial Fusion Energy (IFE) target simulations, Comp. Phys.Comm. 203 (2016) 226, https://doi.org/10.1016/j.cpc.2016.02.014
[14] Atzeni, S., Caruso, A. Inertial confinement fusion: Ignition of isobarically compressed D-T targets. *Nuov Cim B* **80**, 71–103 (1984). https://doi.org/10.1007/BF02899374
[15] Sigmar, D.J., G. Joyce, Plasma heating by energetic particles, Nucl. Fusion 11 (1971) 447, https://doi.org/10.1088/0029-5515/11/5/006

[16] Hora, H. Skin-Depth Theory Explaining Anomalous Picosecond-Terawatt Laser Plasma Interaction II. *Czechoslovak Journal of Physics* 53, 199–217 (2003). https://doi.org/10.1023/A:1022920829925

[17] Lackner, K., Hydrodynamik der Laserfusion, Plenarvortrag Frühjahrstagung Fachverband Atomphysik (1974), in DPG: Vierzig Jahre Fachverband Atomphysik, pg.13

[18] Lackner, K., **"Comments to "A non-thermal laser-driven mixed fuel nuclear fusion reactor concept" by H. Ruhl and G. Korn (Marvel Fusion, Munich)"**, **[v1]** Fri, 1 Apr 2022, **[v2]** Mon, 11 Apr 2022, https://doi.org/10.48550/arXiv.2204.00269

[19] Lackner, K. et al., **"Comments on "Low Q nuclear fusion in a volume heated mixed fuel reactor" by H. Ruhl and G. Korn (Marvel Fusion, Munich)"**, **[v1]** Fri, 17 Mar 2023, https://doi.org/10.48550/arXiv.2303.10017

[20] Lackner, K. et al., "Comments on "Volume ignition of mixed fuel" by H. Ruhl and G. Korn (Marvel Fusion, Munich)", [v1] Tue, 2 May 2023; https://doi.org/10.48550/arXiv.2305.01382

[21] Ruhl, H. and Korn, G., **"Volume ignition of mixed fuel"**, **[v5]** Sat, 25 Feb 2023, https://arxiv.org/abs/2302.06562v5